\documentclass{aa}

\usepackage{graphicx}
\usepackage{txfonts}
\usepackage[pdftitle={Analytic solutions to the maximum and average exoplanet transit depth for common stellar limb darkening laws}, colorlinks = true, breaklinks = true, citecolor = blue, linkcolor = blue, urlcolor = blue, pdfauthor = {Ren\'{e} Heller}]{hyperref}
\usepackage{esvect}  % \vv{}

\begin{document}

\title{Analytic solutions to the maximum and average exoplanet\\ transit depth for common stellar limb darkening laws}
%\subtitle{I. Overviewing the $\kappa$-mechanism}
%\titlerunning{An analytic solution to the maximum exoplanet transit depth for quadratic stellar limb darkening}
\author{Ren\'{e} Heller\inst{1}
          }

   \institute{Max Planck Institute for Solar System Research, Justus-von-Liebig-Weg 3, 37077 G\"ottingen, Germany, \href{mailto:heller@mps.mpg.de}{heller@mps.mpg.de}
             }

   \date{Received 11 November 2018; Accepted 30 January 2019}

% \abstract{}{}{}{}{} 
% 5 {} token are mandatory
 
  \abstract
  {The depth of an exoplanetary transit in the light curve of a distant star is commonly approximated as the squared planet-to-star radius ratio, $(R_{\rm p}/R_{\rm s})^2$. Stellar limb darkening, however, can result in significantly deeper transits. An analytic solution would be worthwhile to illustrate the principles of the problem and predict the actual transit signal required for the planning of transit observations with certain signal-to-noise requirements without the need of computer-based transit simulations.
  }
   {We calculate the overshoot of the mid-transit depth caused by stellar limb darkening compared to the $(R_{\rm p}/R_{\rm s})^2$ estimate for arbitrary transit impact parameters. In turn, this allows us to compute the true planet-to-star radius ratio from the transit depth for a given parameterization of a limb darkening law and for a known transit impact parameter.}
   {We calculated the maximum emerging specific stellar intensity covered by the planet in transit and derive analytic solutions for the transit depth overshoot. Solutions are presented for the linear, quadratic, square-root, logarithmic, and nonlinear stellar limb darkening with arbitrary transit impact parameters. We also derived formulae to calculate the average intensity along the transit chord, which allows us to estimate the actual transit depth (and therefore $R_{\rm p}/R_{\rm s}$) from the mean in-transit flux.}
   {The transit depth overshoot of exoplanets compared to the $(R_{\rm p}/R_{\rm s})^2$ estimate increases from about $15$\,\% for main-sequence stars of spectral type A to roughly $20$\,\% for sun-like stars and some $30$\,\% for K and M stars. The error in our analytical solutions for $R_{\rm p}/R_{\rm s}$ from the small planet approximation is orders of magnitude smaller than the uncertainties arising from typical noise in real light curves and from the uncertain limb darkening.
  }
   {Our equations can be used to predict with high accuracy the expected transit depth of extrasolar planets. The actual planet radius can be calculated from the measured transit depth or from the mean in-transit flux if the stellar limb darkening can be properly parameterized and if the transit impact parameter is known. Light curve fitting is not required.
   }

   \keywords{eclipses -- methods: analytical -- planets and satellites: detection -- stars: atmospheres -- stars: planetary systems -- techniques: photometric}

   \maketitle
%
%-------------------------------------------------------------------

\section{Introduction}
\label{sec:introduction}

The planetary radius ($R_{\rm p}$) is one of the key properties that are currently being derived by several exoplanet hunting surveys. The most successful method to determine the radius of an exoplanet is the transit method, which measures the slight decrease of the brightness of a star as the planet traverses the stellar disk as seen from Earth \citep{1952Obs....72..199S}.

In a simple picture, the star's appearance can be modeled as a circle with uniform brightness. Then the ratio of the planetary radius and stellar radius ($R_{\rm s}$) can be estimated from the constant\footnote{Even with the neglect of stellar limb darkening, however, the transit would still have an ingress and egress, both of which lead to a gradual transition between the in- and out-of-transit apparent stellar brightness \citep[a mathematical description is given in Appendix~A of][]{2014ApJ...787...14H}.} transit depth ($\delta$) via $(R_{\rm p}/R_{\rm s})=\sqrt{\delta}$. In reality, however, stars show center-to-limb brightness variations that affect the estimated planet-to-star radius ratio \citep{2013A&A...549A...9C}.

The common variable used throughout the literature to describe limb darkening is $\mu=\cos(\gamma)$, where $\gamma$ is the angle between the emerging specific intensity of the star and the line of sight of the observer. Early attempts to model stellar limb darkening used equations that are linear in $\mu$ \citep{1921MNRAS..81..361M} but these became insufficient as computer models could be used to model stellar atmospheres. The quadratic \citep{1977A&A....61..809M}, square root \citep{1992A&A...259..227D}, and logarithmic limb darkening laws \citep{1970AJ.....75..175K} provided better agreement with the observations. The advent of exoplanet transit observations \citep{2000ApJ...529L..45C} and new high-accuracy space-based transit photometry \citep{2001ApJ...552..699B}, however, required even better precision. 

\citet{2000A&A...363.1081C} presented a four parameter nonlinear limb darkening law that has now widely been adapted in the exoplanet community in combination with proper limb darkening parameters computed with stellar atmosphere models \citep{2011A&A...529A..75C}. Nevertheless, even though the nonlinear law might be somewhat more precise in certain regimes of the parameter space and for certain stars \citep{2016MNRAS.457.3573E}, the quadratic law is commonly used because \citet{2002ApJ...580L.171M} provided an analytic solution to the resulting transit light curve for the quadratic law.

Although it is well known that the transit depth is not equal to $(R_{\rm p}/R_{\rm s})^2$, no framework exists that conclusively derives the actual deviations from that estimate for stars with limb darkening. Here we derive the correction factors to the $(R_{\rm p}/R_{\rm s})\approx\sqrt{\delta}$ approximation for each of the above-mentioned stellar limb darkening laws.

\section{Methods}
\label{sec:methods}

We present two methods to derive the actual planet-to-star radius ratio for arbitrary parameterizations of the common limb darkening laws. The first method is based on a measurement of the transit depth (Sect.~\ref{sec:overshoot}) and the second method requires a measurement of the arithmetic mean of the in-transit flux (Sect.~\ref{sec:mean}).

%**********************************************
%Fig. 1
\begin{figure}[t]
\centering
\includegraphics[width=1.0\linewidth]{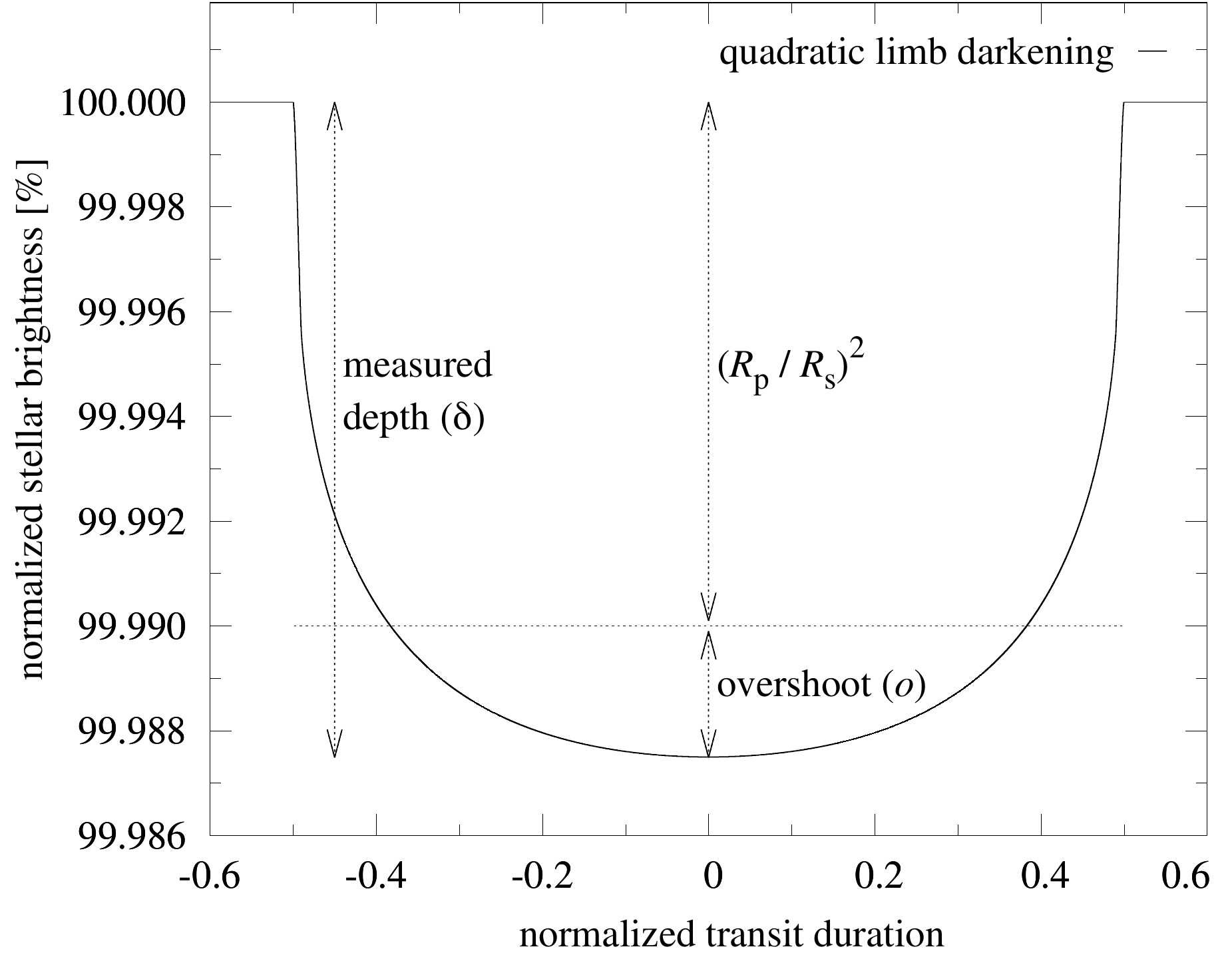}
\caption{Comparison of a simulated transit light curve with quadratic limb darkening \citep[as per][]{2002ApJ...580L.171M} (solid curve) with the $(R_{\rm p}/R_{\rm s})^2$ approximation (dashed line). The planet is assumed to have a radius of 1\,\% the stellar radius, roughly corresponding to an Earth-sized planet around a sun-like star. The stellar limb darkening parameters are arbitrarily set to $a=0.4$ and $b=0.4$ and the transit impact parameter is set to $p=0$. The overshoot indicated by the arrow refers to $o_{\rm LC}$ as defined in Eq.~\eqref{eq:o_LC}.}
\label{fig:LD_depth}
\end{figure}
%**********************************************

\subsection{Transit depth overshoot}
\label{sec:overshoot}

In Fig.~\ref{fig:LD_depth} we show an example transit light curve using the \texttt{python} implementation\footnote{Available at \href{http://www.astro.ucla.edu/\~ianc/files}{http://www.astro.ucla.edu/\textasciitilde{}ianc/files} as transit.py.} of the \cite{2002ApJ...580L.171M} analytic model for quadratic stellar limb darkening by Ian Crossfield. We arbitrarily chose $a=0.4$ and $b=0.4$, irrespective of the stellar spectral type that this choice could imply, a transit impact parameter of zero, and a circular orbit. The solid line shows the numerical computation and the horizontal dashed line shows the $(R_{\rm p}/R_{\rm s})^2$ estimate. The vertical arrow at the bottom of the transit denotes what we refer to as the overshoot ($o$) between the actual transit depth and the $(R_{\rm p}/R_{\rm s})^2$ approximation. There are two ways to calculate $o$ and we can show that they are equivalent to very high precision.

\subsubsection{Transit depth overshoot in the light curve}

The first way to calculate the overshoot uses the light curve. We consider a light curve with an out-of-transit flux normalized to 1, then the transit depth is simply the difference between 1 and the minimum in-transit flux, $\delta=1-f_{\rm min}$. We then define the transit depth overshoot as determined from the light curve as the difference between the transit depth and the $(R_{\rm p}/R_{\rm s})^2$ estimate in units of $(R_{\rm p}/R_{\rm s})^2$,

\begin{equation}\label{eq:o_LC}
o_{\rm LC}= \frac{ \delta - (R_{\rm p}/R_{\rm s})^2 }{(R_{\rm p}/R_{\rm s})^2} = \frac{\delta}{(R_{\rm p}/R_{\rm s})^2} - 1 \ \ .
\end{equation}

\noindent
This is the definition of the overshoot shown in Fig.~\ref{fig:LD_depth} and it illustrates our problem. From an observational perspective, $\delta$ can be readily measured, but $(R_{\rm p}/R_{\rm s})^2$ is what we actually want to determine. Hence, if we knew the transit depth overshoot for a given host star of a transiting planet, then we could correct the simple $(R_{\rm p}/R_{\rm s})\approx\sqrt{\delta}$ approximation and derive the true planet-to-star radius ratio as

\begin{equation}\label{eq:Rp_to_Rs}
{\Bigg (} \frac{R_{\rm p}}{R_{\rm s}} {\Bigg )} = \sqrt{\frac{\delta}{o_{\rm LC}+1}} \ .
\end{equation}

\noindent
Fortunately, there is a second way to calculate the overshoot.

\subsubsection{Overshoot of the stellar limb darkening profile}

This second approach relates to the star's emerging specific intensity, $I$, which we simply refer to as intensity from now on.  The only simplification that we require is to assume that the intensity profile covered by the planet in transit is constant, which is equivalent to the assumption that the planet is infinitesmal. As shown in Sect.~\ref{sec:small}, the resulting error from this small planet approximation is extremely small.

\begin{figure}[t]
\centering
\includegraphics[width=0.775\linewidth]{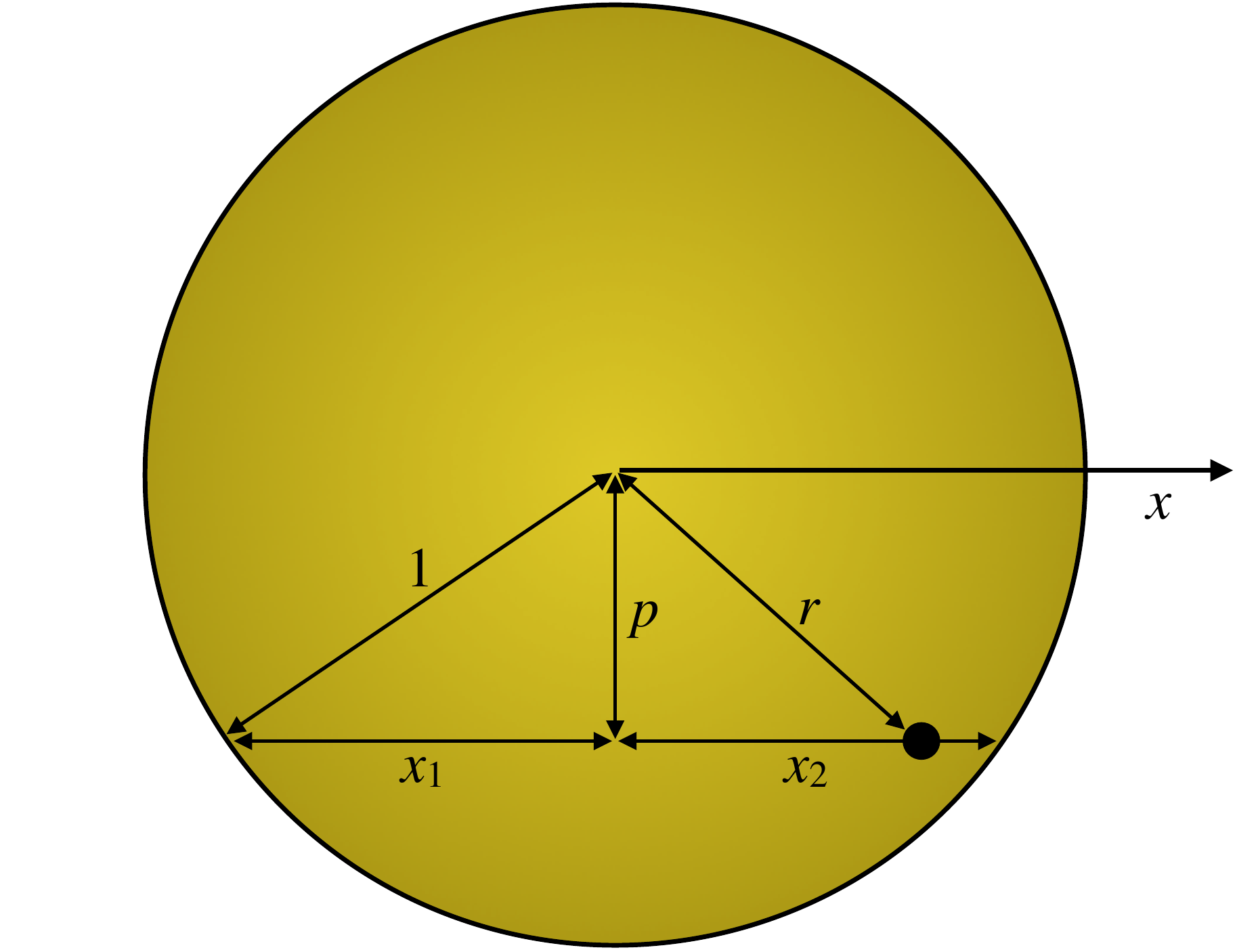}\\
\vspace*{0.3cm}
\includegraphics[width=0.6\linewidth]{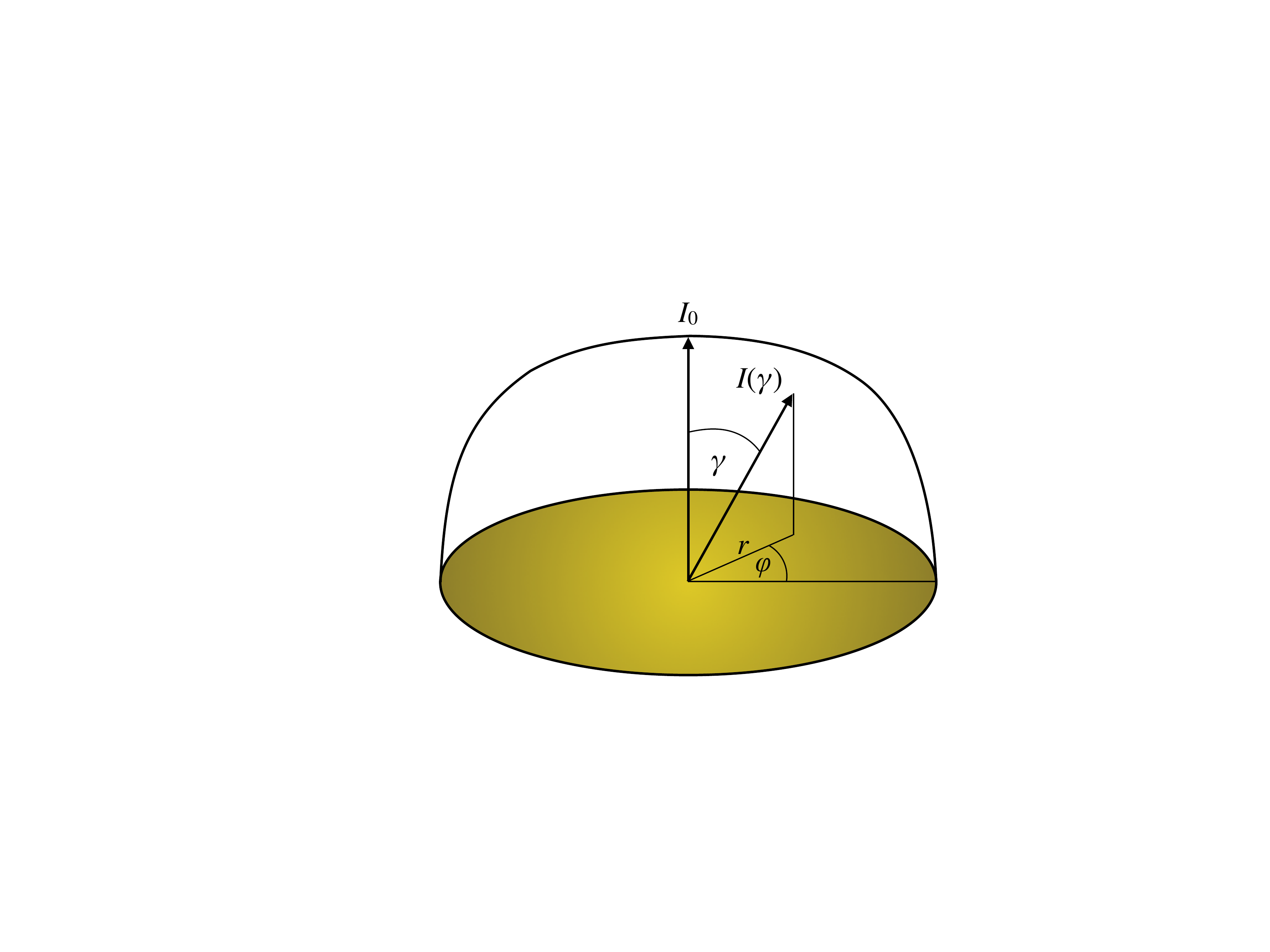}
\caption{{\it Top}: Sketch of the sky-projected apparent stellar disk and illustration of the impact parameter $p$ and integration variables $x$ (along the abscissa) and $r$ (along the radius). The transiting planet is shown as a black circle. {\it Bottom}: Sketch of the stellar intensity or apparent stellar brightness across the disk. The variables $r$ and $\varphi$ are the integration variables used to calculate the disk-averaged intensities for the various limb darkening laws. The curved line represents $I(\gamma)$ in the quadratic limb darkening law with $a=0.4$ and $b=0.4$ as used in Fig.~\ref{fig:LD_depth}.}
\label{fig:LD_sketch}
\end{figure}

We consider a star that is transited by a planet (Fig.~\ref{fig:LD_sketch}). The minimum in-transit distance between the planet and the center of the stellar disk is referred to as the transit impact parameter $0\,{\leq}\,p\,{\leq}\,1$. The radial coordinate, measured from the disk center in units of $R_{\rm s}$, is $r$ and we orient the abscissa, or $x$ coordinate, of our reference system to be parallel to the transit chord of the planet and to have its origin in the center of the stellar disk, so that $r^2=p^2+x^2$. Introducing $x_1$ and $x_2$ as the distances traversed by the planet during the first and second halves of its transit, respectively, we find $1=p^2+x_1^2=p^2+x_2^2$.

The key idea of this study is that the overshoot occurs at the minimum radial distance of the planet to the stellar disk center, that is, when $x=0$ and $r=p$. Without loss of generality, we set the intensity at the disk center $I_0=1$ to simplify our notation. We only consider radially symmetric intensity profiles $I(r)$ in this paper and so we refer to the intensity covered by the planet at mid-transit as $I(r=p)=I_{\rm p}$.

Irrespective of the actual limb darkening law, we can now define the expected transit depth overshoot as the difference between the intensity at mid-transit and the average intensity across the entire stellar disk area, $\langle I \rangle_A$, as

\begin{equation}\label{eq:o_LD}
o_{\rm LD} = \frac{I_{\rm p} - \langle I \rangle_A}{\langle I \rangle_A} = \frac{I_{\rm p}}{\langle I \rangle_A} - 1
\end{equation}

Replacing in Eq.~\eqref{eq:Rp_to_Rs} the overshoot in the light curve $o_{\rm LC}$, which is unknown a priori, with the overshoot of the intensity profile from Eq.~\eqref{eq:o_LD} gives

\begin{equation}\label{eq:Rp_to_Rs_I}
{\Bigg (} \frac{R_{\rm p}}{R_{\rm s}} {\Bigg )} = \sqrt{\delta \ \frac{\langle I \rangle_A}{I_{\rm p}}} \ .
\end{equation}

%**********************************************
%Fig. 2
%**********************************************

\subsection{Mean in-transit flux}
\label{sec:mean}

Alternatively, we can compute the planet-to-star radius ratio from the average in-transit flux ${\langle}f_{\rm in}{\rangle}$ and from the average intensity sampled by the transit chord ${\langle}I{\rangle}_x$. From the mere definition of these two quantities it is clear that

\begin{equation}\label{eq:depth_x}
\delta = \frac{ 1 - {\langle}f_{\rm in}{\rangle} }{{\langle}I{\rangle}_x} \ .
\end{equation}

\noindent
Plugging this expression for $\delta$ into Eq.~\eqref{eq:Rp_to_Rs_I} we obtain

\begin{equation}\label{eq:Rp_to_Rs_x}
\left( \frac{R_{\rm p}}{R_{\rm s}} \right) = \sqrt{ {\Big (}1 - {\langle}f_{\rm in}{\rangle}{\Big )} \frac{ {\langle}I{\rangle}_A }{{\langle}I{\rangle}_x \ I_{\rm p}} } \ ,
\end{equation}

\noindent
which allows us to estimate the actual planet-to-star radius ratio based on the measured mean in-transit flux of a limb-darkened transit light curve and based on the disk average, radial average, and mid-transit intensity of the limb darkening profile.

\subsection{Limb darkening laws}

Now we set out to derive analytical expressions for the average intensity across the stellar disk $\langle I \rangle_A$ and the average intensity along the planetary in-transit path parallel to the $x$ coordinate $\langle I \rangle_x$.

\subsubsection{Linear limb darkening}

We start by rewriting the linear limb darkening law as in \citet{2000A&A...363.1081C}, originally derived by \citep[][Eq.~45 therein]{1921MNRAS..81..361M}, but now as a function of the angle ($\gamma$) between the line of sight and the emerging intensity,

\begin{equation}
I_{\rm lin}(\gamma) = 1 - u \ {\Big (} 1 - \cos(\gamma) {\Big )} \ ,
\end{equation}

\noindent
where $u$ is the limb darkening parameter. With $r=\sin(\gamma)$ as the distance from the star disk center normalized by the stellar radius (see Fig.~\ref{fig:LD_sketch}) and using $\cos\left(\,\arcsin(r)\,\right)~=~\sqrt{1-r^2}$ we have

\begin{equation}\label{eq:I_lin}
I_{\rm lin}(r) = 1 - u \ ( 1 - \sqrt{ 1- r^2 } ) \ .
\end{equation}

\noindent
We then calculate the total intensity across the area elements $({\rm d}\varphi\,{\rm d}r \ r)$ of the stellar disk, where $0\,{\leq}\,\varphi\,{\leq}\,2\pi$ is the azimuth angle, as

\begin{align}\label{eq:I_tot}
I_{\rm tot,lin}  =& \int\displaylimits_{0}^{2\pi} {\rm d}\varphi \int\displaylimits_0^1 {\rm d}r \ r \ I(r)\\ \nonumber
=& \int\displaylimits_{0}^{2\pi} {\rm d}\varphi \int\displaylimits_0^1 {\rm d}r \ r \ {\Big (} 1 - u \ ( 1 - \sqrt{ 1- r^2 } ) {\Big )}\\
=& \ 2\pi \, \frac{3-u}{6} \ .
\end{align}

\noindent
The disk-averaged intensity then is

\begin{equation}\label{eq:I_ave_lin}
\langle I_{\rm lin} \rangle_A = \frac{I_{\rm tot,lin}(u)}{\pi} = 1 - \frac{u}{3} \ ,
\end{equation}

\noindent
where the factor $\pi$ in Eq.~\eqref{eq:I_ave_lin} is the area of the apparent stellar disk with a normalized radius of $r=1$. Using Eq.~\eqref{eq:I_ave_lin} as $\langle I \rangle_A$ and $I_{\rm lin}(r=p)$ from Eq.~\eqref{eq:I_lin} as $I_{\rm p}$, we can estimate the planet-to-star radius ratio from the observed transit depth via Eq.~\eqref{eq:Rp_to_Rs_I} and based on the linear limb darkening law. Alternatively, we can calculate the transit depth overshoot from Eq.~\eqref{eq:o_LD}.

Moving on to the mean intensity across the transit chord, we use the relationship $x_1=\sqrt{1-p^2}=-x_2$ to parameterize the lower and upper boundaries of integration. Instead of integrating the intensity from ingress (at $x_2=-\sqrt{1-p^2}$) to egress (at $x_1=+\sqrt{1-p^2}$), we take advantage of the fact that the average intensity transited by the planet during the first half is the same as during the second half of the transit, which again is equal to the average intensity transited during the entire transit. Hence,

\begin{align}\label{eq:I_ave_lin_x}
{\langle}I_{\rm lin}{\rangle}_x &= \displaystyle \int\displaylimits_{0}^{\sqrt{1-p^2}} {\rm d}x \, I(x) \ {\Big /} \ \displaystyle \int\displaylimits_{0}^{\sqrt{1-p^2}} {\rm d}x \\ \nonumber
&= \frac{1}{\sqrt{1-p^2}} \int\displaylimits_{0}^{\sqrt{1-p^2}} {\rm d}x \, {\Bigg (} 1 - u \ {\Big (} 1 - \sqrt{ 1- (p^2 + x^2) } \, {\Big )} {\Bigg )} \\
&= 1 - u\left(1-\frac{\pi}{4} \sqrt{1-p^2} \right) \ ,
\end{align}

\noindent
which can be used as $\langle I \rangle_x$, for example, in Eq.~\eqref{eq:Rp_to_Rs_x}.

\subsubsection{Quadratic limb darkening}

Rewriting the quadratic limb darkening law as a function of the angle and the normalized radius to the disk center gives

\begin{align}
I_{\rm quad}(\gamma) &= 1 - a\,{\Big (}1-\cos(\gamma){\Big )} - b\,{\Big (}1-\cos(\gamma) {\Big )}^2\\
I_{\rm quad}(r) &= 1 - a\,{\Big (}1- \sqrt{1-r^2} {\Big )} - b\,{\Big (}1 - \sqrt{1-r^2} {\Big )}^2 \ ,
\end{align}

\noindent
where $a$ and $b$ are the two limb darkening parameters. The total disk-integrated intensity $I_{\rm tot,quad}$ can then be calculated as per the right-hand side of Eq.~\eqref{eq:I_tot} and the disk-averaged intensity then turns out as $I_{\rm tot,quad}/\pi$, giving

\begin{equation}\label{eq:I_ave_quad}
\langle I_{\rm quad} \rangle_A = 1 - \frac{a}{3} - \frac{b}{6} \ .
\end{equation}

\noindent
An equivalent expression to Eq.~\eqref{eq:I_ave_quad} has been presented by \citet[][Appendix~B therein]{2013A&A...549A...9C} to calculate the transit depth as a function of $\mu$. \citet{2019A&A...623A..39H} have used this formula for their Transit Least Squares transit detection algorithm ({\tt TLS}).

The transit chord average of the intensity in the quadratic limb darkening law can be obtained in analogy to the previous section using the right-hand side of Eq.~\eqref{eq:I_ave_lin_x}. This results in

\begin{align}\label{eq:I_ave_quad_x}
{\langle}I_{\rm quad}{\rangle}_x &= 1-a\left(1-\frac{\pi}{4}\sqrt{1-p^2}\right)\\
& \hspace{0.3cm} -b\left(\frac{5}{3}-\frac{\pi}{2\sqrt{1-p^2}}-p^2{\Big (}\frac{2}{3}-\frac{\pi}{2\sqrt{1-p^2}}{\Big )}\right) \ .
\end{align}

\subsubsection{Square root limb darkening}
\label{sec:square}

The square root limb darkening law as a function $\gamma$ or $r$ appears as

\begin{align}
I_{\rm sqr}(\gamma) &= 1 - c\,{\Big (}1-\cos(\gamma){\Big )} - d\,{\Big (}1- \sqrt{\cos(\gamma)} {\Big )}\\
I_{\rm sqr}(r) &= 1 - c\,{\Big (}1-\sqrt{1-r^2}{\Big )} - d\,{\Big (}1- \sqrt{\sqrt{1-r^2}} {\Big )} \ ,
\end{align}

\noindent
respectively, where $c$ and $d$ are the corresponding limb darkening parameters. We derive the total disk-integrated intensity $I_{\rm tot,sqr}$ as per Eq.~\eqref{eq:I_tot} and then divide $I_{\rm tot,sqr}$ by the normalized disk area $\pi$ to obtain the disk-averaged intensity as

\begin{equation}\label{eq:I_ave_sqr}
\langle I_{\rm sqr} \rangle_A = 1 - \frac{c}{3} - \frac{d}{5} \ .
\end{equation}

The transit chord average of the intensity in the square root limb darkening then follows from the right-hand side of Eq.~\eqref{eq:I_ave_lin_x},

\begin{align}\label{eq:I_ave_sqr_x}
{\langle}I_{\rm sqr}{\rangle}_x &= 1 - c\left( 1 - \frac{\pi}{4} \sqrt{1-p^2} \right) \\
& \hspace{0.3cm} -d \left( 1 - \frac{\sqrt{\pi} \ (1-p^2)^{1/4} \ \Gamma\left(\frac{5}{4}\right)}{2 \ \Gamma\left(\frac{7}{4}\right)} \right)\ ,
\end{align}

\noindent
with $\Gamma\left(\frac{5}{4}\right)=0.90640...$ and $\Gamma\left(\frac{7}{4}\right)=0.91906...$ (see Appendix~\ref{sec:app_sqr}).

\subsubsection{Logarithmic limb darkening}

The logarithmic limb darkening law can be written as a function of $\gamma$ or $r$ in the following ways,

\begin{align}
I_{\rm log}(\gamma) &= 1 - e\,{\Big (}1-\cos(\gamma){\Big )} - f \cos(\gamma) \, \ln{{\Big(}\cos(\gamma)}{\Big )}\\
I_{\rm log}(r) &= 1 - e\,{\Big (}1-\sqrt{1-r^2}{\Big )} - f \sqrt{1-r^2} \, \ln{{\Big(} \sqrt{1-r^2} {\Big )}} \ ,
\end{align}

\noindent
where $e$ and $f$ are the limb darkening parameters. The total disk-integrated intensity $I_{\rm tot,log}$ then follows via Eq.~\eqref{eq:I_tot} and the disk-averaged intensity is equal to $I_{\rm tot,log}/\pi$, resulting in

\begin{equation}\label{eq:I_ave_log}
\langle I_{\rm log} \rangle_A = 1 + \frac{2}{9}f - \frac{e}{3} \ .
\end{equation}

\noindent
We note that in contrast to all other limb darkening laws treated in this paper, Eq.~\eqref{eq:I_ave_log} can be larger than 1, which means that the star can be brighter on average than in the disk center. In the transit light curve this would manifest itself by a mid-transit depth that is not the deepest point in the transit trough.

The average intensity along the planetary transit chord in the case of logarithmic limb darkening law is computed from the right-hand side of Eq.~\eqref{eq:I_ave_lin_x} as

\begin{align}\label{eq:I_ave_log_x}
{\langle}I_{\rm log}{\rangle}_x = &= 1 - e\left( 1 - \frac{\pi}{4} \sqrt{1-p^2} \right) \\
& \hspace{0.3cm} -f \frac{\pi}{8} \sqrt{1-p^2} \left( 1 - \ln{\Big (}\frac{4}{1-p^2}{\Big )} \right) \ .
\end{align}

\subsubsection{Nonlinear limb darkening}

The nonlinear limb darkening law as a function $\gamma$ and $r$ can be written as

\begin{align}\nonumber
I_{\rm nl}(\gamma) &= 1 - a_1\,{\Big (}1-\sqrt{\cos(\gamma)}{\Big )} - a_2{\Big (} 1-\cos(\gamma) {\Big )} \\ \nonumber
& \hspace{0.62cm} - a_3{\Big (} 1-(\cos(\gamma))^{3/2} {\Big )} - a_4{\Big (} 1 - (\cos(\gamma))^2 {\Big )} \\ \nonumber
I_{\rm nl}(r) &= 1 - a_1{\Big (}1-(1-r^2)^{1/4}{\Big )} - a_2{\Big (}1-(1-r^2)^{1/2}{\Big )} \\
& \hspace{0.62cm}  - a_3{\Big (}1-(1-r^2)^{3/4}{\Big )} - a_4r^2 \ ,
\end{align}

\noindent
with $a_1$ to $a_4$ as the limb darkening parameters. The total disk-integrated intensity for the nonlinear limb darkening law $I_{\rm tot,nl}$ is then obtained from the right-hand side of Eq.~\eqref{eq:I_tot}, and dividing $I_{\rm tot,nl}$ by the normalized apparent disk area $\pi$ gives

\begin{equation}\label{eq:I_ave_nl}
\langle I_{\rm nl} \rangle_A = 1  - \frac{a_1}{5} - \frac{a_2}{3} - \frac{3a_3}{7} - \frac{a_4}{2} \ .
\end{equation}

Finally, radial averaging of the specific intensity in the case of the nonlinear limb darkening law is obtained according to Eq.~\eqref{eq:I_ave_lin_x} as

\begin{align}\label{eq:I_ave_nl_x}\nonumber
{\langle}I_{\rm nl}{\rangle}_x &= 1 - a_1{\Bigg (} 1 - \frac{\sqrt{\pi} \ (1-p^2)^{1/4} \ \Gamma\left(\frac{5}{4}\right)}{2 \ \Gamma\left(\frac{7}{4}\right)}{\Bigg )}\\
& \hspace{0.3cm} -a_2\left( 1 - \frac{\pi}{4} \sqrt{1-p^2} \right)\\
& \hspace{0.3cm} -a_3\left( 1 - \frac{\sqrt{\pi} \ (1-p^2)^{3/4} \ \Gamma\left(\frac{7}{4}\right)}{2 \ \Gamma\left(\frac{9}{4}\right)}\right)\\
& \hspace{0.3cm} \frac{-a_4}{3}(1+2p^2) \ ,
\end{align}

\noindent
where $\Gamma\left(\frac{9}{4}\right)=1.13300...$. This derivation requires solving an integral that does not have a closed form solution, similar to the case of the square root limb darkening law. We nevertheless find the elementary antiderivative shown above by applying a substitution, see the analogous case for the square root limb darkening law in Appendix~\ref{sec:app_sqr}.

\subsubsection{Comparing transit depths between limb darkening laws}

For the linear limb darkening law, we can derive the single limb darkening parameter $u_{o'}$, which produces the same transit depth overshoot as any of the other limb darkening laws. The linear limb darkening law and the quadratic, square, and logarithmic limb darkening laws produce equal overshoots ($o_{\rm LD}=o_{\rm LD}'$) if

\begin{align}
\langle I_{\rm lin} \rangle_A &= \langle I_{\rm quad} \rangle_A \ \ \Leftrightarrow \ \ u_{o'} = a_{o'} + \frac{b_{o'}}{2}\\
\langle I_{\rm lin} \rangle_A &= \langle I_{\rm sqr} \rangle_A \ \ \ \ \Leftrightarrow \ \ u_{o'} = c_{o'} + \frac{3 d_{o'}}{5}\\
\langle I_{\rm lin} \rangle_A &= \langle I_{\rm log} \rangle_A \ \ \ \ \Leftrightarrow \ \ u_{o'} = e_{o'} - \frac{2 f_{o'}}{3}\\
\langle I_{\rm lin} \rangle_A &= \langle I_{\rm nl} \rangle_A \hspace{.48cm} \Leftrightarrow \ \ u_{o'} = \frac{3a_1}{5} + a_2 + \frac{9a_3}{7} + \frac{3a_4}{2} \ .
\end{align}

Of course, the relations between the other limb darkening parameters can be derived analogously by equating the expressions for $\langle I_{\rm quad} \rangle_A$ and $ \langle I_{\rm sqr} \rangle_A$, $\langle I_{\rm quad} \rangle_A$ and $ \langle I_{\rm log} \rangle_A$, and $\langle I_{\rm quad} \rangle_A$ and $ \langle I_{\rm log} \rangle_A$, respectively, and solving for whichever limb darkening parameters are of interest.

\section{Results}
\label{sec:results}

%**********************************************
%Fig. 3
\begin{figure}[t]
\centering
\includegraphics[width=1\linewidth]{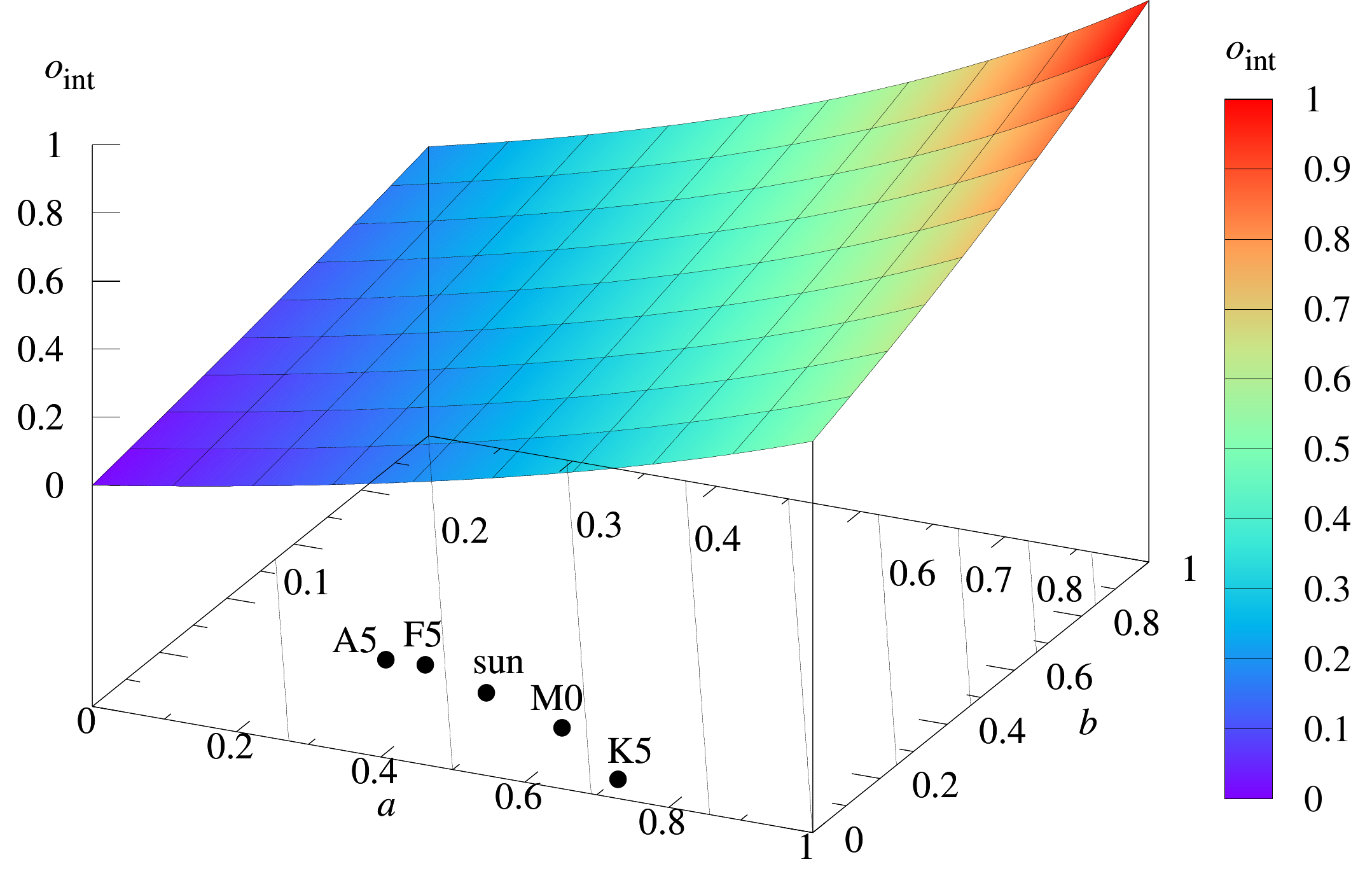}
\caption{Overshoot of the transit depth as a function of the limb darkening parameters $a$ and $b$ in the case of quadratic limb darkening via Eq.~ \eqref{eq:o_LD}. For this plot, a transit impact parameter $p=0$ has been chosen. The positions of several main-sequence stars as per \citet{2011A&A...529A..75C} are indicated with filled circles in the $a$-$b$ plane (see Table~\ref{tab:quadratic}).}
\label{fig:overshoot}
\end{figure}
%**********************************************

\subsection{Transit depth overshoot}

We apply our framework to the numerically computed light curve in Fig.~\ref{fig:LD_depth} as an example. First, we measure $\delta~=~100\,\%~-~99.9875\,\%=0.0125\,\%~=~125$\,parts per million. As this light curve has been computed using the quadratic limb darkening law and using $a=0.4$ and $b=0.4$, we use the same parameterization of Eq.~\eqref{eq:I_ave_quad} and calculate $\langle I_{\rm quad} \rangle_A~=~0.8$. Moreover, with $p=0$ in Fig.~\ref{fig:LD_depth}, we have $I_{\rm p}=1$ and therefore Eq.~\eqref{eq:Rp_to_Rs_I} gives the correct value of $(R_{\rm p}/R_{\rm s})=0.01$ used for this simulation. The corresponding overshoot is calculated from Eq.~\eqref{eq:o_LD} as $o_{\rm LD}=0.25$.

For an analogous application of this procedure to an observed light curve the limb darkening parameters and the transit impact parameter need to be known. The limb darkening parameters can be read from pre-computed tables if the stellar spectral type is known and if the transit impact parameter can be estimated. The latter can be done analytically based on measurements of the transit depth, orbital period, the time between first and fourth contact, and the time between second and third contact \citep{2003ApJ...585.1038S}.

 \begin{table}
\caption{Parameterization of the quadratic limb darkening in the {\it Kepler} band \citep[as per][]{2011A&A...529A..75C} for the main-sequence stars shown in Fig.~\ref{fig:overshoot}. All stars have $\log(g)=4.5$. The corresponding radius overshoot following from Eq.~\eqref{eq:o_LD} is shown in the last column.}              % title of Table
\label{tab:quadratic}      % is used to refer this table in the text
\centering                                      % used for centering table
\begin{tabular}{c c c c c}          % centered columns (4 columns)
\hline\hline                        % inserts double horizontal lines
Spectral Type & $T_{\rm eff}$ & $a$ & $b$ & $o_{\rm LD}$\\    % table heading
\hline                                   % inserts single horizontal line
A5 & 8500 & 0.2474 & 0.3026 & 0.153\\      % inserting body of the table
F5 & 6500 & 0.2987 & 0.3098 & 0.178\\
G2 & 5750 & 0.4089 & 0.2556 &  0.218\\
K5 & 4500 & 0.6809 & 0.0645 & 0.312 \\
M0 & 4000 & 0.5444 & 0.1898 & 0.271\\
\hline                                             %inserts single line
\end{tabular}
\end{table}

Figure \ref{fig:overshoot} illustrates the transit depth overshoot of the quadratic limb darkening law for any combination of $a$ and $b$ and using $p=0$. We use the tables of \citet{2011A&A...529A..75C} to obtain the limb darkening coefficients in the {\it Kepler} band for several main-sequence stars with metallicity [Fe/H]~=~0, surface gravity $\log(g)=4.5$, and zero microturbulent velocity. The spectral types of these stars are indicated as A5, F5, sun, K5, and M0 in Fig.~\ref{fig:overshoot}, respectively. As a result, we find that the quadratic limb darkening law predicts an overshoot of $o_{\rm LD}=0.218$ times $(R_{\rm p}/R_{\rm s})^2$ for a small planet around a sun-like star. This exact solution can be derived by plugging $a$ and $b$ in Eq.~\eqref{eq:I_ave_quad} and the resulting $\langle I_{\rm quad} \rangle_A$ into Eq.~\eqref{eq:o_LD}, where $I_{\rm p}=1$ for a transit impact parameter of zero. Table~\ref{tab:quadratic} lists the overshoots for the other template stars as well, together with the effective temperatures and limb darkening coefficients.

\subsection{Mean in-transit flux}

For the numerically computed example light curve in Fig.~\ref{fig:LD_depth}, where the simulated planet-to-star ratio is 0.01 and the  limb darkening parameters are $a=0.4$ and $b=0.4$ for $\langle I_{\rm quad} \rangle_A$ and $\langle I_{\rm quad} \rangle_x$ in Eqs.~\eqref{eq:I_ave_quad} and \eqref{eq:I_ave_quad_x}, respectively, we measure ${\langle}f_{\rm in}{\rangle}~=~0.999891$. Equation~\eqref{eq:Rp_to_Rs_x} then predicts $(R_{\rm p}/R_{\rm s})~=~0.009996$, the source of the small error in the planetary radius of $4\times10^{-6}\,R_{\rm s}\approx2.8$\,km being in the ingress and egress, as explained in the next section.

\subsection{Small planet approximation}
\label{sec:small}

The error in the planetary radius ($\varepsilon_{\rm p}$) from the small planet approximation can be calculated by comparing the planet-to-star radius ratio $(R_{\rm p}/R_{\rm s})_{\rm obs}$ derived from the measured transit depth in a simulated light curve as per Eq.~\eqref{eq:Rp_to_Rs_I} with the actual planet-to-star radius ratio that went into the simulation, $(R_{\rm p}/R_{\rm s})_{\rm sim}$. The simulation of a Jupiter transit with $p=0$ (hence $I_{\rm p}=1$) across a sun-like star, where $(R_{\rm p}/R_{\rm s})_{\rm sim}=0.099386$ and the quadratic limb darkening parameters are \{$a=0.4089$, $b=0.2556$\}, results in $\delta=1-f_{\rm min}=1.20176\times10^{-2}$. With $\langle I \rangle_A=0.8211$, Eq.~\eqref{eq:Rp_to_Rs_I} then predicts $(R_{\rm p}/R_{\rm s})_{\rm obs}=0.099336$ and the resulting error in the planetary radius of a Jupiter-like planet around a sun-like star is $\varepsilon_{\rm p}=5~{\times}~10^{-5}\,R_\odot\approx35$\,km. For or an Earth-like planet, this same exercise yields an error of $\varepsilon_{\rm p}~=~4~{\times}~10^{-8}\,R_\odot\approx28$\,m.

The main source of error in $(R_{\rm p}/R_{\rm s})_{\rm obs}$ as predicted from transit depth is in the variation of the intensity over the small area of the stellar disk that is temporarily occulted by the planet. The larger the planet, the larger the variation, and the larger the error.

For both of these test cases, we also calculate the mean in-transit flux ${\langle}f_{\rm in}{\rangle}$ to predict $(R_{\rm p}/R_{\rm s})_{\rm obs}$ via Eq.~\eqref{eq:Rp_to_Rs_x}. For the Jupiter-sized test planet, we measure ${\langle}f_{\rm in}{\rangle}=0.9894685$ and with $\langle I \rangle_A=~0.8211$ and $\langle I \rangle_x=0.887745$ I obtain $(R_{\rm p}/R_{\rm s})_{\rm obs}~=~0.0986959$. The resulting deviation from the injected planet-to-star radius ratio is $\varepsilon_{\rm p}=6.9~{\times}~10^{-4}\,R_\odot\approx480$\,km. The same exercise for the Earth-like planet yields an error of $\varepsilon_{\delta}=4.2\times10^{-6}\,R_\odot\approx2.9$\,km.

The main source of error in $R_{\rm p}/R_{\rm s}$ as predicted from the mean flux is in the ingress and egress of the planet, which is not taken into account in our calculations of the radially averaged emerging specific intensity of the star. The larger the planet, the longer the times of ingress and egress, and the larger the error in the resulting $(R_{\rm p}/R_{\rm s})_{\rm obs}$ estimate from the chord-averaged intensity.

These tests show that the effect from the small planet approximation is orders of magnitude smaller than any uncertainties arising from the total noise budget in real light curves. Even high-accuracy space-based observations of the brightest and most photometrically quiet stars \citep{2011ApJS..197....6G} have a noise floor of several times 10 parts per million, which translates into an error in the derived radius ratio of $>10^{-4}$ for $R_{\rm p}/R_{\rm s}=0.1$ and an error $>10^{-3}$ for $R_{\rm p}/R_{\rm s}=0.01$ (the relative error increases for smaller planets). Our small planet approximation produces errors in $R_{\rm p}/R_{\rm s}$ that are orders of magnitude smaller than the error coming from the uncertain limb darkening coefficients \citep{2013A&A...560A.112M}; this is a crucial issue for any fitting of observed light curves with numerical methods.

\section{Conclusions}

In this paper we present analytical expressions for the overshoot of the transit depth for small planets ($R_{\rm p}~{\ll}~R_{\rm s}$) with arbitrary transit impact parameter for the linear, quadratic, square-root, logarithmic, and nonlinear stellar limb darkening laws. Equation~\eqref{eq:o_LD} can be used to calculate the overshoot of the stellar emerging intensity with respect to the disk-averaged intensity, which translates into an overshoot of the observed transit depth. Equation~\eqref{eq:Rp_to_Rs_I} gives the actual planet-to-star radius ratio based on the measured transit transit depth ($\delta$) and based on the limb darkening law and parameterization. For sun-like stars, exoplanet transits can be $\sim20$\,\% deeper than the $(R_{\rm p}/R_{\rm s})^{2}$ estimate.

We also derive analytical expressions to calculate the maximum transit depth $\delta$ from the mean in-transit flux ${\langle}f_{\rm in}{\rangle}$ for any of the limb darkening laws mentioned above. This approach is based on calculating the average stellar intensity along the transit chord of the planet across the stellar disk ${\langle}I{\rangle}_x$, which relates to ${\langle}f_{\rm in}{\rangle}$ and $\delta$ as per Eq.~\eqref{eq:depth_x}. Combined with our results for the overshoot of the transit depth, we derive Eq.~\eqref{eq:Rp_to_Rs_x} which gives an estimate for the actual planet-to-star radius ratio based only on the measured in-transit mean flux and on the parameterization of the respective limb darkening law, which requires knowledge of the transit impact parameter.

These expressions can be used to calculate the expected maximum transit depth for a given planet-star system with any parameterization of the common stellar limb darkening laws. This approach is not meant to replace the numerical fitting of light curves, the latter of which can constrain limb darkening and transit impact parameter. Our approach is certainly less accurate than the use of elliptical integrals \citep{2013MNRAS.432.2216A} or than the full modeling of the stellar intensity as a sum of spherical harmonics \citep{2019AJ....157...64L}, let alone the numerical simulation of transits with model stellar atmospheres \citep{2018arXiv180502696N}. But our approach requires no numerical modeling and it could be more intuitive to understand.

\begin{acknowledgements}
The author is thankful to Michael Hippke for posing the problem solved in this study and to an anonymous referee for her/his useful comments. This work was supported by the German space agency (Deutsches Zentrum f\"ur Luft- und Raumfahrt) under PLATO Data Center grant 50OO1501. This work made use of NASA's ADS Bibliographic Services
\end{acknowledgements}

\bibliographystyle{aa}
\bibliography{ms}

\onecolumn

\begin{appendix}

\section{Average intensity of the transit chord for the square root limb darkening law}
\label{sec:app_sqr}

In Sect.~\ref{sec:square}, we present our result for the average intensity sampled by the planet along its transit chord in the case of the square root limb darkening law. Its derivation in analogy to Eq.~\eqref{eq:I_ave_lin_x} is a bit more complicated than for the other limb darkening laws because the corresponding integral $\int {\rm d}x \, I_{\rm sqr}(x)$ does not have a closed form integral, hence we need to work on that. First, we separate the integral into its summands,

\begin{align}
{\langle}I_{\rm sqr}{\rangle}_x &= \displaystyle \int\displaylimits_{0}^{\sqrt{1-p^2}} {\rm d}x \, I_{\rm sqr}(x) \ {\Big /} \ \displaystyle \int\displaylimits_{0}^{\sqrt{1-p^2}} {\rm d}x = \frac{1}{\sqrt{1-p^2}} \int\displaylimits_{0}^{\sqrt{1-p^2}} {\rm d}x \, {\Bigg (} 1 - c \ {\Big (} 1 - \sqrt{ 1- (p^2 + x^2) } \, {\Big )} {\Bigg )} - d \ {\Big (} 1 - \left( 1- (p^2 + x^2) \right)^{1/4} \, {\Big )} {\Bigg )} \\
&= \frac{1}{\sqrt{1-p^2}} {\Bigg [} \int\displaylimits_{0}^{\sqrt{1-p^2}} {\rm d}x \, 1 \ - \ c \int\displaylimits_{0}^{\sqrt{1-p^2}} {\rm d}x \, \left( 1 - \sqrt{1-(p^2+x^2)} \ \right) - \ d \int\displaylimits_{0}^{\sqrt{1-p^2}} {\rm d}x \, \left( 1 - \left(1-(p^2+x^2)\right)^{1/4} \right) \ {\Bigg ]} \\
&= \frac{1}{\sqrt{1-p^2}} {\Bigg [} \sqrt{1-p^2} - c \ {\Big (} \sqrt{1-p^2} - \frac{\pi}{4}(1-p^2) {\Big)} - \ d \int\displaylimits_{0}^{\sqrt{1-p^2}} {\rm d}x \, 1 + \ d \int\displaylimits_{0}^{\sqrt{1-p^2}} {\rm d}x \, \left(1-(p^2+x^2)\right)^{1/4} \ {\Bigg ]} \\ \label{eq:I_sqrt_long}
&= \frac{1}{\sqrt{1-p^2}} {\Bigg [} \sqrt{1-p^2} - c \ {\Big (} \sqrt{1-p^2} - \frac{\pi}{4}(1-p^2) {\Big)} - \ d \sqrt{1-p^2} + \ d \int\displaylimits_{0}^{\sqrt{1-p^2}} {\rm d}x \, \underbrace{ \left(1-p^2-x^2)\right)^{1/4}}_{\displaystyle \hspace{0.87cm} \equiv (u)^{1/4}} \ {\Bigg ]} \ ,
\end{align}

\noindent
and we identify the last summand as the integral that does not have an elementary antiderivative. Hence, we substitute $1-p^2-x^2\,{\equiv}\,u$, which means

\begin{align}
x = \sqrt{1-p^2-u} \ \ , \ \ \frac{{\rm d}x}{{\rm d}u} = \frac{-1}{2\sqrt{1-p^2-u}} \ \ , \ \ u(x=0) = 1-p^2 \ \ , \ \ u(x=\sqrt{1-p^2})=0 \ ,
\end{align}

\noindent
so that

\begin{align}
\int\displaylimits_{0}^{\sqrt{1-p^2}} {\rm d}x \, \left(1-p^2-x^2)\right)^{1/4} = -\int\displaylimits_{1-p^2}^{0} {\rm d}u \, \left(-\frac{1}{2} \frac{1}{\sqrt{1-p^2-u}})\right) \ u^{1/4} \ = \ + \frac{1}{2} \int\displaylimits_{0}^{1-p^2} {\rm d}u \, \frac{u^{1/4}}{\sqrt{1-p^2-u}} = \frac{\sqrt{\pi} \ (1-p^2)^{3/4} \ \Gamma\left(\frac{5}{4}\right)}{2 \ \Gamma\left(\frac{7}{4}\right)} \ .
\end{align}

\noindent
Equation~\eqref{eq:I_sqrt_long} then simplifies to

\begin{align}
{\langle}I_{\rm sqr}{\rangle}_x &= 1 - c\left( 1 - \frac{\pi}{4} \sqrt{1-p^2} \right) -d \left( 1 - \frac{\sqrt{\pi} \ (1-p^2)^{1/4} \ \Gamma\left(\frac{5}{4}\right)}{2 \ \Gamma\left(\frac{7}{4}\right)} \right)\ ,
\end{align}

\noindent
which is Eq.~\eqref{eq:I_ave_sqr_x}.

\end{appendix}

\end{document}